# Comment for article "Limits on a nucleon-nucleon monopole-dipole (axionlike) P-,T-noninvariant interaction from spin relaxation of polarized ultracold neutrons" Yu.N. Pokotilovski, arXiv:0902.3425v2 [nucl-ex] 22 Feb 2009.

A.P. Serebrov

*Petersburg Nuclear Physics Institute RAS*

**Abstract**

In work [1] (Yu. N. Pokotilovski, arXiv:0902.3425v2) restrictions on constants of pseudo-magnetic interaction $g_S g_P$ are presented. These restrictions are considerably differed from restrictions on $g_S g_P$, before published in work [2] (A.P. Serebrov, arXiv:0902.1056v1). Restrictions in work [1] are received from the same experimental data which are used in work [2], however difference in restrictions is considerable. This difference is changed in a range from 1 to $10^7$ times depending on value $\lambda$. In the given work it is shown that restrictions of work [1] are wrong and the possible reasons of the admitted errors are considered.

Let's consider in more details the task about UCN depolarization at reflection from walls of UCN storage trap (Fig. 1). UCN depolarization arises due to a pseudo-magnetic field near to vertical walls of the tarp, because a pseudo-magnetic field direction is orthogonal to a leading vertical magnetic field $H_z$. The UCN depolarization effect at one wall collision can be calculated in system of coordinates of a moving neutron and in rotating system of coordinates. The frequency of rotating system of coordinates has to be equal to neutron spin Larmor frequency round a magnetic field $H_z$. In this system of coordinates the magnetic field $H_z$ appears completely compensated, and the pseudo-magnetic field becomes variable:

$$H(t) = H_r(t)\cos\omega_z t, \quad (1)$$

where $\omega_z = 2\pi\gamma H_z$, $\gamma$ - neutron gyromagnetic ratio.

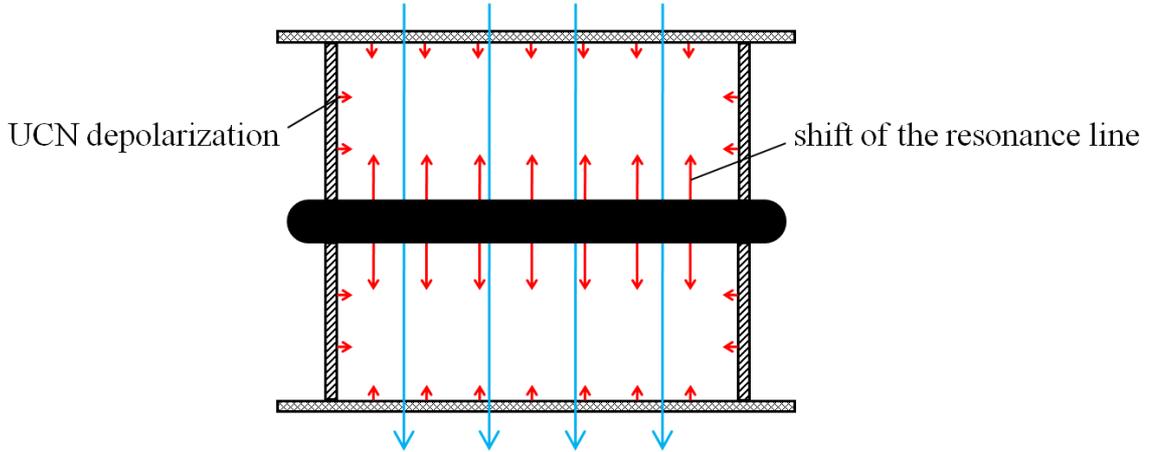

Fig. 1. Scheme of the experiment for neutron electric dipole moment search. The pseudo-magnetic neutron spin precession in the vicinity of vertical walls leads to a random neutron spin flip and UCN depolarization during their storage. The pseudo-magnetic neutron spin precession in the vicinity of horizontal walls will cause the neutron resonance shift, if the central and external electrodes are made from materials of different densities.

The pseudo-magnetic field near to a surface is described by dependence [2]:

$$H(r) = H_{r_0}(\lambda)e^{-|r-r_0|/\lambda}, \quad (2)$$

where $H_{r_0} = \dfrac{\hbar^2 \lambda}{4m_n\mu_n} N g_S g_P$ (see [2]).

In system of coordinates of a moving neutron $H(r)$ is transformed to $H_r(t)$:

$$H_r(t) = H_{r_0}(\lambda)e^{-|t|/\tau_\lambda}, \quad (3)$$

where $\tau_\lambda = \lambda / v_n$, $v_n$ - normal component of speed to a wall surface.

Thus, in rotating system of coordinates of a moving neutron $H(t) = H_{r_0}(\lambda)e^{-|t|/\tau_\lambda}\cos\omega_z t$.

It is necessary to calculate depolarization effect at one wall collision. Polarization on $z$ axis interests us. $P_z = P_0 \cos\theta$, where $\theta$ - deviation angle of $P_0$ from $z$ axis after wall collision. At small angles $\theta$: $P_z = P_0(1 - \theta^2/2)$. Depolarization effect ($\beta$) is equal to $\theta^2/2$.

$$\theta = 2\pi\gamma H_{r_0} \int_{-\infty}^{+\infty} e^{-|t|/\tau_\lambda}\cos\omega_z t\, dt = \frac{4\pi\gamma H_{r_0}\tau_\lambda}{1+(\omega_z\tau_\lambda)^2} \quad (4)$$

$$\theta = \frac{2\omega_\lambda \tau_\lambda}{1+\left(\omega_z \tau_\lambda\right)^2}, \qquad (5)$$

where $\omega_\lambda = 2\pi\gamma H_{r_0}(\lambda)$.

$$\beta = \frac{1}{2}\left[\frac{2\omega_\lambda \tau_\lambda}{1+(\omega_z \tau_\lambda)^2}\right]^2 \Bigg|_{(\omega_z \tau_\lambda)^2 \ll 1} \approx \frac{1}{2}\left(2\omega_\lambda \tau_\lambda\right)^2 \qquad (6)$$

In work [2] the case of $(\omega_z \tau_\lambda)^2 \ll 1$ is considered. It is a condition of non-adiabatic pseudo-magnetic field occurrence because time of action of a pseudo-magnetic field $2\tau_\lambda$ is much less than rotation period of spin round a magnetic field $H_z$.

When $(\omega_z \tau_\lambda)^2 \gg 1$, it is a case of adiabatic pseudo-magnetic field occurrence because during pseudo-magnetic field action $2\tau_\lambda$ there are many turns round a magnetic field $H_z$ and the depolarization effect is suppressed.

$$\beta = \frac{1}{2}\left[\frac{2\omega_\lambda \tau_\lambda}{1+(\omega_z \tau_\lambda)^2}\right]^2 \Bigg|_{(\omega_z \tau_\lambda)^2 \gg 1} \approx \frac{1}{2}\left[2\frac{\omega_\lambda \tau_\lambda}{(\omega_z \tau_\lambda)^2}\right]^2 \qquad (7)$$

In work [1] the second case is wrongly chosen, because instead of $\tau_\lambda$ it was considered $\tau_c$ (time between neutron collisions with walls). Besides, in work [1] condition of adiabaticity is accepted a priori on a condition $H_\lambda / H_z \ll 1$. It is necessary, but not a sufficient condition of adiabaticity. As a result, in work [1] the formula for adiabatic case is applied though actually the case is non-adiabatic $(\omega_z \tau_\lambda)^2 \ll 1$. The small parameter $\omega_z \tau_\lambda$ appears in a power of 4 in a denominator of the formula (7). It leads to an error of big orders.

Besides, from the formula (7) follows that the depolarization probability does not depend from $\lambda$ because $\omega_\lambda$ and $\tau_\lambda$ are proportional $\lambda$. This erroneous conclusion of independence from $\lambda$ is transferred on restrictions on $g_S g_P$ as it is seen in Fig. 2. (Fig. 1 from work [1].) In Fig. 2 constraints from work [2] correspond to curve 3, constraints from work [1] correspond to curve 6.

Certainly, at reduction of $\lambda$ divergence degree between formulas (6) and (7) increases reaching 7 orders of magnitude. ($\omega_z \tau_\lambda$ becomes ever less, but adiabatic case is used.)

For example, it is easy to calculate deviation angle of vector $P_0$ from $z$ axis for $\lambda = 10^{-4}$ cm and $g_S g_P = 10^{-20}$ (an extreme point at the left in Fig. 2):

$$\theta = \frac{2\omega_\lambda \tau_\lambda}{1+\left(\omega_z \tau_\lambda\right)^2} \approx 2\omega_\lambda \tau_\lambda = 2.2 \cdot 10^{-10} \text{ rad.}$$

Accordingly, the depolarization effect ($\beta = \theta^2/2 = 2.5 \cdot 10^{-20}$) is less on 15th orders of magnitude than experimental value $\beta_{\exp} = 10^{-5}$, but in work [1] experimental value $10^{-5}$ was used to obtain constraints for $g_S g_P = 10^{-20}$ at $\lambda = 10^{-4}$ cm. This estimation is the obvious proof of abnormality of restrictions on size of $g_S g_P$ in work [1].

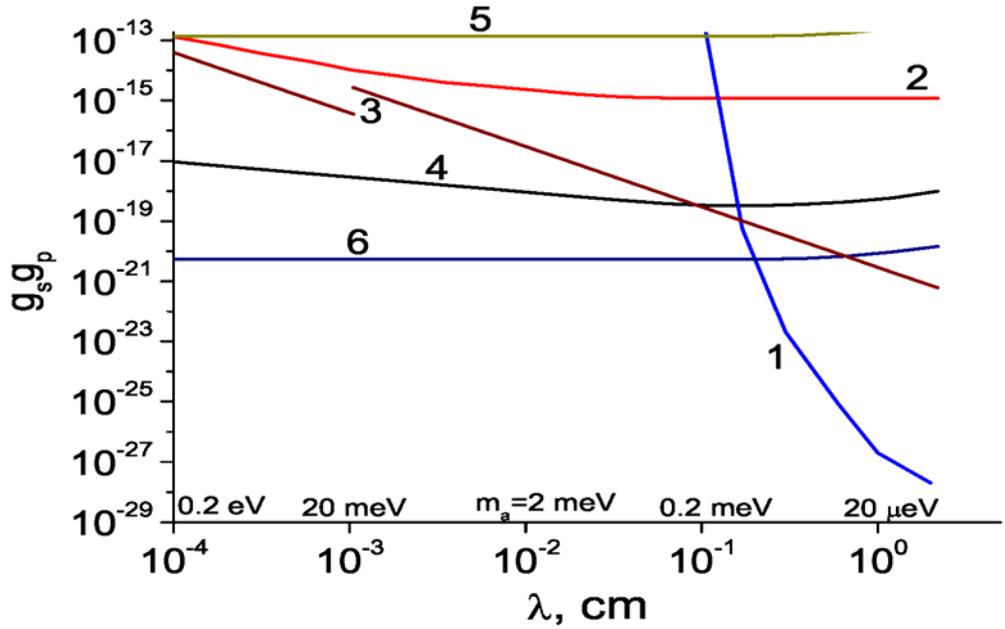

Fig. 2. Constraints on the axion monopole-dipole coupling strength $g_S g_P$ and effective range: 1 - from Ref. [4], 2 - from Ref. [5], 3 - from Ref. [2], 4 - from spin relaxation of $^3$He, Ref. [3], 5 – work [1] in an assumption that the UCN depolarization probability $\beta = 10^{-5}$ and magnetic field $H_z = 50$ G [6,7], 6 - the same, but $H_z = 0.01$ G [8,9].

In summary it is necessary to notice that in work [3] ("Limits on a nucleon-nucleon monopole-dipole axionlike P-,T-noninvariant interaction from spin relaxation of polarized He-3", Yu.N. Pokotilovski, arXiv:0902.1682v2) the same formulas, as in work [1] were applied. It raises the big doubts in justice of conclusions of work [3] also.

**References**

[1] Yu.N. Pokotilovski, arXiv:0902.3425v2.
[2] A.P. Serebrov, arXiv:0902.1056v1.
[3] Yu.N. Pokotilovski, arXiv:0902.1682v2.
[4] R.C. Ritter, L.I. Winkler, and G.T. Gillies, Phys. Rev. Lett., 70 (1993) 701;
 Wei-Tou Ni, Shean-Shi Pan, sien-Chi Yeh, Li-Shing Hou, and Juling Wan, Phys. Rev. Lett., 82 (1999) 2439.
[5] S. Baeßler, V.V. Nesvizhevsky, K.V. Protasov, and A.Yu. Voronin, Phys. Rev. D75 (2007) 075006.
[6] A.P. Serebrov, M.S. Lasakov, A. Vasilijev, I.A. Krasnoshchokova, Yu.A. Rudnev, A. Fomin, V.E. Varlamov, P. Geltenbort, J. Butterworth, A.R. Young, and U. Pasavento, Nucl. Instr. Meth., A440 (2000) 715, and Phys. Lett., A313 (2003) 373.
[7] F. Atchison, B. Brys, M. Daum, P. Fierlinger, P. Geltenbort, R. Henneck, S. Heule, M. Kasprzak, K. Kirch, K. Korch, A. Pichlmaier, Ch. Plonka, U. Straumann, C. Wermelinger, and G. Zsigmond, Phys. Rev., C76 (2007) 044001.
[8] I.S. Altarev, Yu.V. Borisov, N.V. Borovokova, A.I. Egorov, S.N. Ivanov, E.A. Kolomensky, M.S. Lasakov, V.M. Lobashev, V.A. Nazarenko, A.N. Pirozhkov, A.P. Serebrov, Yu.V. Sobolev, E.V. Shulgina, Yad. Fiz., 59 (1996) 1204; Phys. At. Nucl., 59 (1996) 1152.
[9] C.A. Baker, D.D. Doyle, P. Geltenbort, K. Green, M.G.D. van der Grinten, P.G. Harris, P. Iaydjiev, S.N. Ivanov, P.J.R. May, J.M. Pendlebury, J.P. Richardson, D. Shiers, and K.F. Smith, Phys. Rev. Lett., 97 (2006) 131801.